\documentclass[pra,a4paper,twocolumn,showpacs]{revtex4}
\usepackage{graphicx}
\newcommand{\be}{\begin{equation}}
\newcommand{\ee}{\end{equation}}
\newcommand{\bea}{\begin{eqnarray}}
\newcommand{\eea}{\end{eqnarray}}
\newcommand{\ba}{\begin{array}}
\newcommand{\ea}{\end{array}}
\newcommand{\ben}{\begin{enumerate}}
\newcommand{\een}{\end{enumerate}}
\newcommand{\bei}{\begin{itemize}}
\newcommand{\eei}{\end{itemize}}

\newcommand{\p}{\partial}

\newcommand{\om}{\omega}

\newcommand{\bd}{b^\dagger}
\newcommand{\cd}{c^\dagger}
\newcommand{\vd}{v^\dagger}

\newcommand{\dek}{\Delta\epsilon_k}

\newcommand{\dnk}{\Delta n_k}

\newcommand{\la}{\langle}
\newcommand{\ra}{\rangle}
\newcommand{\pd}{P^\dagger}

\newcommand{\lm}{\lambda}
\newcommand{\lmq}{{\lambda q}}
\begin{document}
\title{Semiconductor Microstructure in a Squeezed Vacuum: \\ Electron-Hole Plasma Luminescence}
\author{Eran Ginossar and Shimon Levit}
\affiliation{Department of Condensed Matter Physics, The Weizmann Institute of Science, Rehovot 76100, Israel}
\email{eran.ginossar@weizmann.ac.il}
\date{\today}
\begin{abstract}
We consider a semiconductor quantum-well placed in a wave guide microcavity and interacting with the broadband
squeezed vacuum radiation, which fills one mode of the wave guide with a large average occupation. The wave guide
modifies the optical density of states so that the quantum well interacts mostly with  the squeezed vacuum.
The vacuum is squeezed around the externally controlled central frequency $\om_0$, which is tuned above the electron-hole gap $E_g$,
and induces fluctuations in the interband polarization of the quantum-well.
The power spectrum of scattered light exhibits a peak around $\om_0$, which is moreover non-Lorentzian and is  a result of both the
squeezing and the particle-hole continuum. The  squeezing spectrum is  qualitatively different  from
the atomic case. We discuss the possibility to observe the above phenomena
in the presence of additional non-radiative (e-e,\,\,phonon) dephasing.
\end{abstract}

\maketitle
\section{Introduction}
The modification of the spontaneous emission of an atom placed in a cavity has been predicted and observed
a long time ago \cite{Kleppner81,goy}.
Recently, cavity effects have been observed also in quantum dots and quantum wells \cite{Solomon} which were placed in
 a microcavity made of Distributed Bragg Mirrors (DBRs). Similarly, effects of non-classical radiation, such as squeezed vacuum states
produced in the process of parametric down-conversion, were  considered in the context of interaction
 with atoms \cite{Gardiner86,Palma-Knight,Charmichael87}.
An important  theoretical prediction  was made by Gardiner \cite{Gardiner86},
that an atom coupled to a squeezed reservoir
will exhibit two line widths in its resonance fluorescence spectrum. To observe this effect, it is necessary that most of the
electromagnetic modes which are resonant with the atomic transition are occupied by squeezed vacuum with a
large average photon occupation.

In this letter we consider the coupling of a two-band system of a semiconductor quantum well to a squeezed
reservoir of photons occupying the modes of an ideal optical wave guide (without leakage).
This presents a generalization of the atomic case in several ways: (1) there is a continuum of electron-hole excitations
in the band, (2) each particle-hole excitation is detuned differently with respect to the squeezing energy $\om_0$ (which
is externally controlled),
(3) inevitable additional nonradiative relaxation and dephasing.
The solid-state environment involves more types of interactions which have to be considered together,
but it offers a possibility to observe these quantum optical effects, as the quality of cavities improves.
Specifically, we consider the luminescence of the scattered squeezed state, in the regime where the nonradiative
dephasing ($\gamma_2$) is of the order or smaller than the radiative transition rate ($\Gamma$), and the
relaxation ($\gamma_1$) of electrons is small compared to $\Gamma$. In this regime, where the radiation
may be assumed to be a reservoir, we will argue that the effect of Coulomb interaction is mainly to give rise to dephasing.
We calculate the optical spectra of the {\em unshifted luminescence}, i.e. the scattered radiation with the same
frequency as the incoming radiation and compare it qualitatively to the atomic case.
The power spectrum (Fig. \ref{peak-figure}) has a non-Lorentzian peak at the frequency $\om_0$, a consequence of the strong energy
dependence of the correlation times of the fluctuating polarization of different e-h pairs.
The squeezing spectrum exhibits reduced fluctuations in
one quadrature (Fig. \ref{squeeze-figure}), the minimum of which is proportional to $\sqrt{\Gamma+\gamma_2-|\Lambda|}$,
where $\Lambda$ is a measure of the squeezing correlations of the field.

\section{Model}
We will first present our model system and the results for the
scattered radiation, and later discuss its possible realization.
The free part of the Hamiltonian $H=H_0+H_I+H_c$ is given by ($\hbar\equiv 1$)
\be\label{H0_ham}
H_0=\sum_\lmq\omega_\lmq\bd_\lmq b_\lmq +\sum_k\epsilon_k^c\cd_kc_k+\sum_k\epsilon_k^v\vd_kv_k. \ee
The operators $c_k$ and $v_k$ denote annihilation operators of the free electrons in the conduction and valence
 bands of the quantum well, with in-plane momentum $k$ (omitting spin).
The operators $b_\lmq$ denote the photon annihilation operators of the wave guide mode $\lm$ with wave number $q$.
For simplicity we confine ourselves to the case of normal incidence, i.e. $q$ is orthogonal
to the electronic in-plane momentum $k$.
The interaction of the electrons and the photons in the dipole approximation is given by \cite{Haug-Koch-Book,Khitrova-Review}
\be\label{V_int}
 H_I=\sum_{k \lm q} \mathcal{E}(\om_\lmq)u_\lm b_\lmq(d_{cv}\cd_kv_k+d^*_{cv}\vd_kc_k)+\,\,\,h.c. \ee
where $\mathcal{E}(\om_\lmq)=\sqrt{\frac{\om_\lmq}{V}}$, $u_\lm$ is an overlap integral of the electronic envelope and wave guide mode functions,
 $d_{cv}$ is dipole matrix element (we suppress its weak dependence on $k$
\footnote{The dependence of the dipole matrix element on $k$ can be neglected since we are interested only in the luminescence
in the band $\om_0\pm B/2$ where $B$ is much smaller than the electronic band.}), and $V$ is the wave guide volume.
Finally, the Coulomb interaction is given by \cite{Haug-Koch-Book}
\bea \label{coul-ham}
&& H_c=\frac{1}{2}\sum_{k,k',q\neq 0} U_q \left(2\cd_{k+q}\vd_{k'-q}v_{k'}c_k + \right.\\ \nonumber
&& \left. +\cd_{k+q}\cd_{k'-q}c_{k'}c_k+\vd_{k+q}\vd_{k'-q}v_{k'}v_k \right)
\eea
where $U_q$ is the bare Coulomb interaction.

We assume that the radiation acts as a reservoir with  correlations of a two mode broadband squeezed vacuum \cite{Gardiner86}
\bea \label{field-corrs}
&&\langle \bd_\lmq b_{\lm' q'}\rangle=\mathcal{N}(\om_\lmq)\delta_{\lm\lm'}\delta(\om_\lmq-\om_{\lm q'}) \\ \nonumber
&&\langle b_\lmq b_{\lm' q'}\rangle=\mathcal{M}(\om_\lmq)\delta_{\lm\lm'}\delta(2\om_0-\om_\lmq-\om_{\lm q'}) \\ \nonumber
&&\langle b_\lmq \rangle =0\\ \nonumber
\eea
where $\mathcal{N}(\om_\lmq)$ is the average photon occupation of the mode $\lm$ and wave number $q$, and $\mathcal{M}(\om_\lmq)$ is the squeezing parameter \cite{Gardiner86}.
We assume that $\mathcal{N},\, \mathcal{M} \gg 1$ within the bandwidth $\om_0\pm B/2$ of
one {\em squeezed} mode (denoted by $\lm=s$) and they vanish for other {\em empty} modes (denoted by $\lm=e$).

The energy of the central frequency $\om_0$ is tuned higher than the electron-hole gap $E_g$, and the bandwidth of the squeezed
radiation ($B$) is assumed to be much larger than the radiative width ($\Gamma$), but much smaller than the conduction
  band width. The radiation induces interband transitions between the valence and conduction bands.
In principle, when the photon occupation is large, it is possible
to find the system in a regime where the dephasing rates due to
non-radiative and radiative scattering are comparable, and
nonradiative relaxation is small. In this regime the system,
initially an unexcited full valence band, reaches a stationary
state in which there is a large occupation of the conduction band
in the energy stripe where the photon occupation is large.

In order to understand qualitatively the effect of the squeezed reservoir on the electron-hole pair,
it is instructive to begin with the equation of motion for $\la\pd_k(t)\ra=\la \cd_k(t)v_k(t)\ra$,
the interband polarization, without taking into account explicitly the $H_c$ part of the Hamiltonian.
Using the Markov approximation \cite{Haug-Koch-Book} we find in the  rotating frame
of frequency $\dek=\epsilon^c_k-\epsilon^v_k$,
\be \label{eom-for-p}
\frac{d\langle\pd_k(t)\rangle}{dt}=-(\Gamma+\gamma_2)\langle\pd_k(t)\rangle+e^{2i\delta_k t}\Lambda\langle P_{k}(t)\rangle
\ee
where $\delta_k=\om_0-\dek$ is the detuning frequency, $\gamma_2$ is the dephasing rate due to electron-electron
 and electron-phonon scattering. The radiative decay coefficients $\Gamma$ and $\Lambda$ are given
(for the squeezed mode $\lm=s$) by
\bea \label{decay-coefs}
&& \Gamma(\dek)=\rho(\dek)|A_s(\dek)|^2(\mathcal{N}(\dek)+\frac{1}{2}) \\ \nonumber
&& \Lambda(\dek)=\rho(\dek)\mathcal{M}(\dek)A_s^*(\dek)A_s^*(2\om_0-\dek) \\ \nonumber
\eea
where $\rho$ is the optical density of states, and $A_s(\dek)=\mathcal{E}(\dek)u_s d_{cv}$ is the electron-photon
coupling strength. It is assumed that there is only a small number of empty resonant modes, thus neglecting
their influence in the dynamics of $\la\pd_k\ra$.

The special features of equation (\ref{eom-for-p}) are: (1) the coupling of $\pd_k$ to  $P_k$,
due to the squeezing correlations (Eq. (\ref{field-corrs})) in the radiation, and (2) the absence of a coupling to the
population inversion $\dnk=n^v_k-n^c_k$, due to the zero average of the field, Eq. (\ref{field-corrs}).
The solution of (\ref{eom-for-p}) exhibits two complex frequencies
\be \Omega_{1,2}(k)=i(\Gamma+\gamma_2)+\delta_k \mp i\sqrt{|\Lambda|^2-\delta_k^2}.\ee
For $\delta_k<|\Lambda|$, the frequencies $\Omega_{1,2}(k)$ have two different imaginary
parts, which can be interpreted as {\em over-damping} of the polarization,
 whereas for $\delta_k>|\Lambda|$ the $\Omega_{1,2}(k)$ have two different real parts,
 reminiscent of {\em under-damping}.

The incoming squeezed vacuum radiation can not induce average interband polarization.
However, it  induces fluctuations $\la \pd_k(t+\tau) P_{k'}(t)\ra$ and $\la P_k(t+\tau) P_{k'}(t)\ra$ of the polarization, which
in turn give rise to scattered radiation in all the resonant wave guide modes. These correlators can be measured by power and squeezing spectra.
First we calculate the power spectrum of the scattered radiation into an initially empty mode of the wave guide
($\lm=e$).
In the long-time limit it is given by
\bea \label{lum-integ-orig}
&&\mathcal{L}(\om_{\lmq})\equiv\frac{\partial}{\partial t}N_{e,q}(t)=\sum_{kk'}|A_e(\om_{e,q})|^2\times \\ \nonumber
&&\times\int_{-\infty}^{\infty}d\tau e^{-i(\om_{e,q}-\dek)\tau}
\la\pd_k(t+\tau)P_{k'}(t)\ra
\eea
where $N_{e,q}$ is the occupation of the photon mode $\lm=e$ with wave number $q$.

Let us begin by considering the Coulomb interaction. The effective mean field Hamiltonian \cite{lindberg-koch} which usually leads to coherent
excitonic correlations in the e-h plasma is given by
\bea
&& H_{eff}=-\sum_{k,q\neq 0}U_q\left(n^c_{k-q}\cd_kc_k+n^v_{k-q}\vd_kv_k+ \right. \\ \nonumber
&& \left. +p^*_{k-q}\vd_kc_k+p_{k-q}\cd_kv_k \right)
\eea
where $n^{c,v}_k$ are band occupations and $p_k=\la P_k \ra$.
For the dynamics of $\la \pd_k(t+\tau) P_{k'}(t)\ra$ w.r.t. $\tau$, under the condition $p_k=0$ this interaction
 can only induce energy level shifts.
Going beyond the mean field approximation it is simplest to consider the equation of motion for the
correlations $C_{kk'}\equiv\la \pd_k(t) P_{k'}(t)\ra$.
This equation contains contributions of the form  $(1-n^c_k-n^v_{k-q})\sum_{k'}U_{k'-k}C_{kk'}$, which in principle give rise
to an excitonic effect \footnote{This issue has been recently debated in several theoretical and experimental works, see
for example \cite{kira-hoyer-koch, chatterjee} }.
In the presence of the photon reservoir which we consider here, we assume
the band occupations $n^{c,v}_k$ to be close to $1/2$ (see below), leading to a reduction of the excitonic effect through the
phase filling prefactor $(1-n^c_k-n^v_{k-q})$. As a result of the above considerations we will assume from now on that
the effect of Coulomb interaction is twofold. First it gives rise to energy shifts which
we assume constant over the range $B$, and include in $\epsilon^{(v,c)}_k$. Second, it induces
dephasing which we assume is included in phenomenological constant $\gamma_2$.

Let us turn next to the interaction of the electronic band with the squeezed reservoir.
The equation of motion for $\la \pd_k(t+\tau) P_{k'}(t)\ra$ with respect to $\tau$ can be derived from the
explicit equation of motion for the operator $\pd_k(t)$ (see appendix), which a form similar to Eq. (\ref{eom-for-p}).
The solution for $\tau>0$ is given by
\bea\label{fluc-sol}
&&\la \pd_k(t+\tau)P_{k'}(t)\ra=\\ \nonumber
&& =G_k(\tau)\la \pd_k(t)P_{k'}(t)\ra+H_k(\tau)e^{2i\delta_k t}\la P_k(t)P_{k'}(t)\ra
\eea
where
\bea \nonumber \label{fluc-funcs}
&& G_k(\tau)=\frac{1}{2}\left[e^{i\Omega_1(k)\tau}(1-i\frac{\delta_k}{\Delta_k})+e^{i\Omega_2(k)\tau}(1+i\frac{\delta_k}{\Delta_k})\right] \\
&& H_k(\tau)=\frac{\Lambda}{2\Delta_k}(e^{i\Omega_1(k)\tau}-e^{i\Omega_2(k)\tau}).
\eea
where $\Delta_k=\sqrt{|\Lambda|^2-\delta_k^2}$. For $\tau<0$ the solution is given by taking $G^*_k(-\tau)$ and $H^*_k(-\tau)$.
 The dependence of the last term on $t$ in Eq. (\ref{fluc-sol}) is due to the non-stationary nature of the squeezed radiation \cite{Gardiner86}.

The off diagonal $k\neq k'$ elements of equal time correlations
$\la \pd_k(t) P_{k'}(t)\ra$ and $\la P_k(t) P_{k'}(t)\ra$ which serve as initial conditions in (\ref{fluc-sol}) vanish in equilibrium.
However here the  system is coupled to two reservoirs: the nonradiative thermal bath and the squeezed reservoir.
It can be shown \cite{unpublished} that as a result there are non-zero steady state off-diagonal correlations
of two kinds: {\em normal} $\la \pd_k(t) P_{k'}(t)\ra$ and {\em anomalous} $\la P_k(t) P_{k'}(t)\ra$. They
are smaller than  the diagonal correlations $\la \pd_k(t) P_{k}(t)\ra$
by a factor proportional to $(\frac{\gamma_1}{\Gamma})^2$.  Qualitatively this can be estimated
by referring to the equation $\p_tP_k = d_{cv} \sum_\lmq \mathcal{E}(\om_\lmq)u_\lm b_\lmq\Delta n_k$.
The off diagonal correlations  are driven by the squeezed reservoir since
there exists a non-zero steady state difference $\la\Delta n_k\ra$.
This is given  by the rate equation
\be \p_t \Delta n_k(t)=-(\Gamma+\gamma_1)\Delta n_k(t)+\gamma_1\ee
which leads to the above estimate.

\section{Results}
We shall treat the luminescence in the limit  of $\gamma_1/\Gamma \to 0$ so that only the
diagonal correlations $\la \pd_k(t+\tau) P_{k}(t)\ra$
contribute to (\ref{lum-integ-orig}). Moreover, since $\la P_k(t) P_k(t)\ra=0$,
and $\la \pd_k(t) P_k(t)\ra=\la n^c_k\ra\la 1-n^v_k\ra$,   Eq. (\ref{fluc-sol}) shows that
$\la \pd_k(t+\tau) P_{k}(t)\ra$ are stationary in this limit.

\begin{figure}
\includegraphics[scale=0.58]{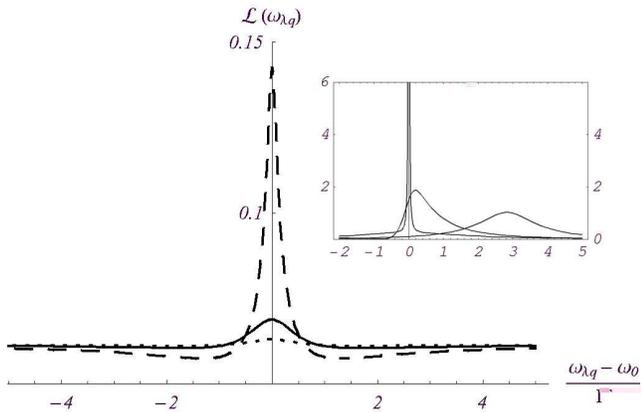}
\caption{Unshifted luminescence ($\mathcal{L}$) for squeezing $\Lambda=0.9\Gamma$, bandwidth $B=50\Gamma$, and  phonon dephasing
 $\gamma_2=0$ ({\em dashed curve}), $\gamma_2=3\Gamma$ ({\em dotted curve}), and $\gamma_2=\Gamma$ ({\em solid curve}). Insert: particle-hole lineshapes
for different detuning $\delta_k=0,0.9\Gamma$ and $3\Gamma$ (from left to right). The curves are normalized to the total power.}
\label{peak-figure}\end{figure}

Substituting (\ref{fluc-sol}) with $k=k'$ and (\ref{fluc-funcs}) in the integral (\ref{lum-integ-orig}) we obtain
\begin{widetext}
\be\label{lum-integ}
\mathcal{L}(\om_\lmq)= 2|A_\lm(\om_{\lmq})|^2Re\int d^2k\la n^c_k\ra\la 1-n^v_k\ra
\left[\frac{i}{\dek-\om_\lmq+\Omega_1(k)}(1-i\frac{\delta_k}{\Delta_k})+\frac{i}{\dek-\om_\lmq+\Omega_2(k)}(1+i\frac{\delta_k}{\Delta_k})\right]\ee
\end{widetext}
where we assume $\lm=e$ from now on. The lineshape of an individual  particle-hole transition (the integrand of
(\ref{lum-integ})) has two distinct limits (see insert of figure (\ref{peak-figure})): when $\delta_k \rightarrow 0$ it consists of two superimposed regular
Lorentzian lineshapes of different widths $\Gamma+\gamma_2\pm|\Lambda|$.
When $\delta_k\gg\Lambda$ the lineshape is a single Lorentzian of width $\Gamma+\gamma_2$, which is just the radiative width
acquired without squeezing. In between those limits the lineshapes are asymmetric with short tails lying on the
side of the central frequency $\om_0$. The superposition of such lineshapes produces a non-Lorentzian peak at  $\om_0$ in
the unshifted luminescence spectrum $\mathcal{L}(\om_\lmq)$, Fig. (\ref{peak-figure}).
For simplicity  $\la n_k\ra$, $\mathcal{N}(\dek)$, $A_\lm(\om_\lmq)$ and $\mathcal{M}(\dek)$ were
taken as constant in the energy range $\om_0\pm B/2$ (these will be assumed also below for the squeezing
spectrum).

The width of the peak is of the order of $\Gamma+\gamma_2$, and the maximum is (in the $B\gg \Gamma$ limit)
\be \mathcal{L}_{max}=\mathcal{L}_0 \frac{\Gamma +\gamma_2}{\sqrt{(\Gamma+\gamma_2)^2-|\Lambda|^2}},\ee
where $\mathcal{L}_0=4\pi^2|A_\lm(\om_\lmq)|^2\rho_{el}\la n^c_k\ra\la 1-n^v_k\ra$ with  $\rho_{el}$ the electronic density of states.

\begin{figure}
\includegraphics[scale=0.76]{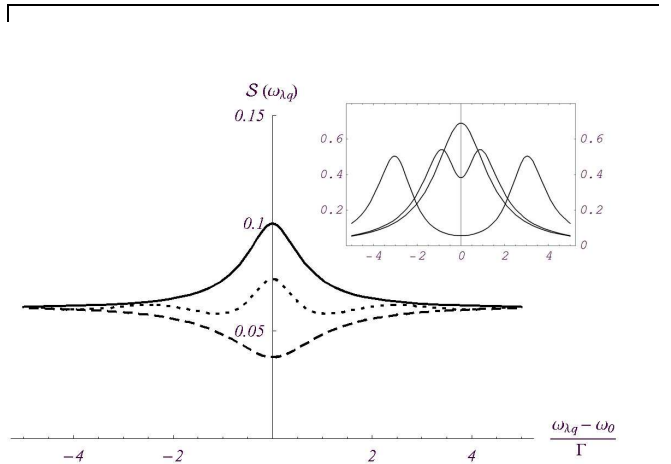}
\caption{Squeezing spectra ($\mathcal{S}(\om_\lmq)$) for squeezing $\Lambda=0.9\Gamma$, bandwidth $B=50\Gamma$, phonon dephasing
of $\gamma_2=\Gamma$. It shows the in-phase quadrature fluctuations ({\em solid curve}) and out-of-phase quadrature
fluctuations ({\em dashed curve}) for a fixed squeezing angle ($\alpha=0$). The linearly modulated squeezing phase spectrum is given
for the in-phase quadrature ({\em dotted curve}). Insert: particle-hole squeezing spectra
for different detuning $\delta_k=0,0.9\Gamma$ and $3\Gamma$ (from left to right) in the out-of-phase quadrature.}
\label{squeeze-figure}
\end{figure}

We now turn to the squeezing spectrum which is defined   \cite{Gardiner-Parkins-Collet} in terms of the fluctuations of the field  quadrature
 $X_\theta=\frac{1}{2}\left[ E^{(+)}(t)e^{i\om_0t-i\theta}+E^{(-)}(t)e^{-i\om_0t+i\theta}\right]$,
\be \mathcal{S}_{\theta}(\om_\lmq)=\int_{-\infty}^{\infty}d\tau \la : X_\theta(t+\tau)X_\theta(t) :\ra e^{i\Delta\om_\lmq\tau}\ee
where $\Delta\om_\lmq\equiv \om_\lmq-\om_0$ and $\theta$ determines the choice of quadrature.
For the limit $\gamma_1\ll \Gamma$ we again neglect the off-diagonal correlations $\la P_k(t)P_{k'}(t)\ra$ ($k\neq k'$)
and obtain in the long time limit,
\begin{widetext}
\bea\label{sq-integ}
&& \mathcal{S}_{\theta}(\om_\lmq)=\frac{1}{2} \psi^2 Re\int d^2k\la n^c_k\ra\la 1-n^v_k\ra \left[ \left(1+\frac{|\Lambda|\cos(2\theta+
\alpha)}{\Delta_k}\right)\frac{-1}{i\Delta\om_\lmq+\Delta_k-(\Gamma+\gamma_2)}+ \right. \\ \nonumber
&&+\left.\left(1-\frac{|\Lambda|\cos(2\theta+
\alpha)}{\Delta_k}\right)\frac{-1}{i\Delta\om_\lmq-\Delta_k-(\Gamma+\gamma_2)}\right]\eea
\end{widetext}
where $\alpha$ is the phase of $\Lambda$, Eq. (\ref{decay-coefs}), and $\psi$ is a geometrical factor
\footnote{We absorb a constant phase of $d_{cv}$ in the definition of $\theta$. $\psi$ depends on the mode
$\lambda$, the parameters of the waveguide, and on the position of the detector \cite{carmichael-book}.}.
Figure (\ref{squeeze-figure}) shows the squeezing spectra for the in-phase  ($2\theta+\alpha=0$) and out-of-phase
($2\theta+\alpha=\pi$) quadratures, normalized to the total emitted power. The minimum of the out-of-phase quadrature
is proportional to $\sqrt{\frac{\Gamma+\gamma_2-|\Lambda|}{\Gamma+\gamma_2+|\Lambda|}}$, and thus can in principle
 be very small for ideal squeezing ($\Lambda \rightarrow \Gamma$) and vanishing non-radiative dephasing ($\gamma_2\rightarrow 0$).
Note that the squeezing spectrum is defined with respect to the
{\em normally ordered} correlation function so that a zero value means reduction to vacuum fluctuations.
The squeezing spectrum of a band is qualitatively different from that of a single
particle-hole (insert of Fig. (\ref{squeeze-figure})), which has a maximum proportional to $\frac{1}{\Gamma+\gamma_2+|\Lambda|}$ at $\om_\lmq-\om_0\approx \delta$
(for strong squeezing $\Lambda\approx \Gamma$).
 Note also that the squeezing phase $\alpha$ need not be constant and
can be modulated as a function of  the frequency.
For example, a linear dependence $\alpha=2|\delta_k|/\Gamma$ produces oscillations in the squeezing spectrum
(Fig. (\ref{squeeze-figure})).

\section{Realization}
\begin{figure}
\includegraphics[scale=0.4]{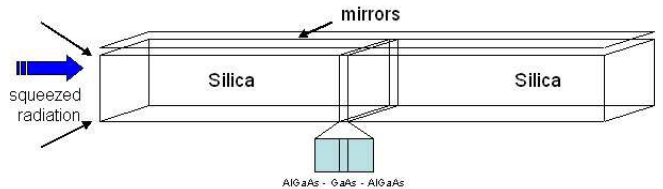}
\caption{Schematic realization of the quantum well embedded in a wave guide cavity.}
\label{waveguide-fig}
\end{figure}

A possible realization of the above model is shown in Fig. (\ref{waveguide-fig}) with the wave guide
made of, e.g. metallic or Bragg mirrors \footnote{Modification of spontaneous emission in a
wave guide was considered for example in \cite{Kleppner81} and in a micro-wave guide in the context of semiconductors
in \cite{brorson}.}.
To minimize optical losses in the wave guide, which reduce the squeezing ($\Lambda$)
of the light \cite{artoni-loudon,schmidt,patra}, the interior of the wave guide should be filled with a material
such as silica whose absorption is negligible.
The reduction of squeezing due to losses is exponential \cite{artoni-loudon}, with an exponent $-4\alpha l$, where
$\alpha$ is the absorption coefficient, and $l$ is the distance.
The length of the wave guide ($L$) should be much larger than the typical wavelength $\lambda=c/\om_0$ to ensure that coupling transients in the
vicinity of the opening do not reach the quantum well. Therefore $L$ should be of the order of $100-1000\mu m$ at least.
The effect of dispersion is to rotate the squeezing phase \cite{blow-loudon}, since pairs of correlated photon modes acquire
a relative fixed phase over the distance in the dispersive material. This has no consequence for the power spectrum,
but may result in a modulation of the squeezing phase with frequency, which can modify the squeezing spectrum, as was
discussed above.

The central frequency $\om_0$ should be  above the cutoff frequency of the lowest wave guide mode.
The range ($\om_0\pm B/2$) should be tuned above the exciton ionization energy
and below the optical phonon energy.
The latter should make it possible to avoid the fast emission of optical phonons.
The temperature should be small \cite{Levinson} compared to $\sqrt{(\hbar\om_0-E_g) m^*s^2}$ ($s$ is the velocity of sound),
 so that spontaneous emission of acoustic phonons is the dominant electron-phonon scattering mechanism.
The other important non-radiative source of dephasing is Coulomb scattering of the photo-excited electrons. This scattering
rate is proportional to the density, and hence in our case to the bandwidth, for measurement times shorter than $\gamma_1^{-1}$.

We estimated the non-radiative rates for InAs \cite{esipov} using  the momentum relaxation time
as an estimate for $\gamma_2$. In the quasi-elastic regime the momentum relaxation time is of the same order of
the scattering rate \cite{Levinson}. A simple Golden Rule calculation gives for a Boltzmann gas
$\frac{1}{\tau_{e-e}}=\pi^2\frac{\hbar\epsilon_B}{m^*\epsilon}N$, where $N$ is surface density and $\epsilon_B=me^4/2\hbar^2$.
For a zero lattice temperature and taking $\om_0=E_g+20meV$,
$\Gamma=1\mu eV$, $B=10\mu eV$ and density $2\times 10^8\, cm^{-2}$ we found this estimate  to be $3.3\mu eV$ (i.e. $200ps$).
These estimates are roughly supported experimentally \cite{Wang, Haacke}. In these experiments
the dephasing times for the interband polarization were measured directly, and were shown to be of the order of $5ps$ for
excitation density $N\sim 10^{10}\,\,\, cm^{-2}$, a density which is two order of magnitude higher than the one we consider ($N\sim 10^8\,\,\, cm^{-2}$).
The dephasing times were also shown in one of the experiments \cite{Haacke} to be much longer for the lower densities.
The remaining dephasing in those densities was attributed to disorder, which we believe is a much weaker scattering
process in the electron-hole plasma regime.

\section{Conclusions}
To conclude, we found non-trivial lineshape structures in the spectra of a two mode squeezed vacuum scattered off a
two band electronic system in two dimensions. They
appear despite the flat spectrum of the incoming radiation and constant electronic density  of states.
These effects are due to the particular correlations  of the squeezed vacuum and seem to be experimentally observable.
\section{Acknowledgements}
It is a pleasure to acknowledge valuable discussions with Y. B. Levinson and J. G. Groshaus.
This work is supported by DIP Grant No. DIP-C7.1.\\

\appendix

\section*{Appendix: The Polarization Fluctuation Equation}

Here we derive the effective equations of motion which lead to the solution (\ref{fluc-sol}) for
the fluctuations. We employ a consistent truncation of
the hierarchy of equations of motion at the level of three and four-body correlations. We further assume
 the previous conditions $\Gamma\gg \gamma_1$.
The e.o.m. for $\la \pd_k(t)P_{k'}(t')\ra$ w.r.t. to one time argument is derived from the Hamiltonian $H=H_0+H_I$
\bea \label{fluc-eom}
&& \partial_t \la \pd_k(t)P_{k'}(t')\ra=i\dek\la \pd_k(t)P_{k'}(t')\ra+ \\ \nonumber
&& +i\sum_qA^*_q\la\bd_q(t)\dnk(t)P_{k'}(t')\ra .\eea
The derivative of $\la\bd_q(t)\dnk(t)P_{k'}(t')\ra$ is given by the two terms
\bea  \nonumber && \la\dot{\bd}_q(t)\dnk(t)P_{k'}(t')\ra=i\om_q\la\bd_q(t)\dnk(t)P_{k'}(t')\ra+\\
&&\label{bnp-eom-1} +iA_q\sum_{k''}\la \pd_{k''}(t)\dnk(t)P_{k'}(t')\ra \\ \nonumber
&&  \label{bnp-eom-2}\la\bd_q(t)\dot{\dnk}(t)P_{k'}(t')\ra=2i\sum_{q'}A_{q'}\la\bd_q(t)b_{q'}(t)\pd_k(t)P_{k'}(t')\ra- \\
&& -2i\sum_{q'}A^*_{q'}\la\bd_q(t)\bd_{q'}(t)P_k(t)P_{k'}(t')\ra .\eea
We now assume that $\la \pd_{k''}(t)\dnk(t)P_{k'}(t')\ra\approx \la \pd_{k''}(t)P_{k'}(t')\ra\la\dnk(t)\ra$ is very small, since in the steady state
$\la\dnk(t)\ra\ll 1$ (note that the correlation $\la \pd_{k''}(t)\dnk(t)P_{k'}(t')\ra$ cannot decouple
to contributions with an average polarization $\la\pd_k\ra$).
Next we assume that at the second order in the system-reservoir interaction $A_q$, the correlations can
be decoupled by approximating for the total density matrix $\rho_{tot}=\rho_{sys}\otimes\rho_{res}$.
As a result we get for the correlations in the r.h.s. of Eq. (\ref{bnp-eom-2})
\bea \nonumber && \la\bd_q(t)b_{q'}(t)\pd_k(t)P_{k'}(t')\ra \approx \la\bd_q(t)b_{q'}(t)\ra\la\pd_k(t)P_{k'}(t')\ra \\ \nonumber
&& \la\bd_q(t)\bd_{q'}(t)P_k(t)P_{k'}(t')\ra \approx \la\bd_q(t)\bd_{q'}(t)\ra\la P_k(t)P_{k'}(t')\ra .\nonumber \eea
Substituting all the contributions in (\ref{bnp-eom-2}) back into (\ref{fluc-eom}) we have for $\la \pd_k(t)P_{k'}(t')\ra$
\begin{widetext}
\bea
&&  \partial_t \la \pd_k(t)P_{k'}(t')\ra=i\dek\la \pd_k(t)P_{k'}(t')\ra
-2\sum_{qq'} A_q^* A_{q'}e^{i\om_q t} \int^t dt'' e^{-i\om_q t''}\la\bd_q(t'')b_{q'}(t'')\ra\la\pd_k(t'')P_{k'}(t')\ra+  \\ \nonumber
&& +2\sum_{qq'} A_q^* A_{q'}^* e^{i\om_q t}\int^t dt''e^{-i\om_q t''}\la\bd_q(t'')\bd_{q'}(t'')\ra\la P_k(t'')P_{k'}(t')\ra.\eea
Similarly, the equation for the correlation $\la P_k(t)P_{k'}(t')\ra$ is given by
\bea
&&  \partial_t \la P_k(t)P_{k'}(t')\ra=-i\dek\la P_k(t)P_{k'}(t')\ra-
2\sum_{qq'} A_q A_{q'}^*e^{-i\om_q t} \int^t dt'' e^{i\om_q t''}\la\bd_q(t'')b_{q'}(t'')\ra\la\pd_k(t'')P_{k'}(t')\ra+  \\ \nonumber
&& +2\sum_{qq'} A_q A_{q'} e^{-i\om_q t}\int^t dt''e^{i\om_q t''}\la b_q(t'')b_{q'}(t'')\ra\la P_k(t'')P_{k'}(t')\ra.\eea

We now employ the correlations (\ref{field-corrs}) and the Markov approximation to obtain

\bea
&& \partial_t \la \pd_k(t)P_{k'}(t')\ra=(i\dek-\Gamma(\dek))\la \pd_k(t)P_{k'}(t')\ra+\Lambda(\dek)e^{2i\delta_k t}\la P_k(t)P_{k'}(t')\ra \\ \nonumber
&& \partial_t \la P_k(t)P_{k'}(t')\ra=(-i\dek-\Gamma(\dek))\la P_k(t)P_{k'}(t')\ra+\Lambda^*(\dek)e^{-2i\delta_k t}\la \pd_k(t)P_{k'}(t')\ra. \eea
\end{widetext}

\end{document}